\documentstyle{article}
\def\be{\begin{eqnarray}}
\def\ee{\end{eqnarray}}
\def\nn{\nonumber}

\begin{document}

\titlepage

\phantom. \hfill ITEP-M1/94 \\

\bigskip

\bigskip

\centerline{\Large NEW MATRIX MODEL SOLUTIONS}
\centerline{\Large TO THE KAC-SCHWARZ PROBLEM}

\bigskip

\bigskip

\centerline{M.Adler}
\centerline{\it Brandeis University, Waltham, Mass 02254, USA}

\bigskip

\centerline{A.Morozov}
\centerline{\it ITEP, 117259, Moscow, Russia}

\bigskip

\centerline{T.Shiota}
\centerline{\it Dep.of Mathematics, Kyoto University,
Kyoto 606, Tokyo}

\bigskip

\centerline{P.van Moerbeke}
\centerline{\it Universite de Louvain, 1348 Louvain-la-Neuve, Belgium}
\centerline{\it Brandeis University, Waltham, Mass 02254, USA}

\bigskip

\bigskip

\bigskip

\centerline{ABSTRACT}

\bigskip

We examine the Kac-Schwarz problem of specification of
point in Grassmannian in the restricted case of
gap-one first-order differential Kac-Schwarz operators.
While the pair of constraints satisfying $[{\cal K}_1,W] = 1$
always leads to Kontsevich type models, in the case of
$[{\cal K}_1,W] = W$ the
corresponding KP $\tau$-functions are represented as
more sophisticated matrix integrals.

\newpage
\section{Introduction}

In the framework of string theory one is interested, among other
things, in construction of the non-perturbative partition function
of a string model. By definition this ``non-perturbative partition
function'' $\tau\{t\}$ is a generating functional for all the
correlators in the given model and as such, it depends on some set
of parameters $\{t\}$, in which it can be expanded as a formal
serius. It also depends on particular model, i.e. on some other
set of parameters $\{M\}$, parametrizing the ``module space of
models''. Since ``exponentiation of perturbations'' - implicit
in the concept of generating functional - effectively changes the
action, i.e. parameters of the model, there are all reasons to
believe that in fact the nature of both types of parameters
is almost the same, and ``universal partition function''
$\tau\{t|M\}$ essentially depends on some specific ``combination''
$t\circ M$ of variables $t$ and $M$:
\be
\tau\{t|M\} \sim \tau\{t\circ M\}.
\label{combin}
\ee

So far little is known about $\tau\{t|M\}$ in such a general situation
\cite{AM}.
Particular example, studied already in some detail, is
provided by the theory of Generalized Kontsevich Model (GKM), i.e.
the family of matrix models, closely related to that of the
simplest Landau-Ginzburg topological gravity. This set of models
is parametrized by a function $W(x)$ and universality classes
are labeled by types of singularities of $W(x)$. In the most
popular case $W(x)$ is a polinomial of degree $p$, $W(x) = W_p(x)$.
Then the variables $t$ form a sequence (i.e. a discrete one-parametric
set, with discreteness reflecting
the absense of particle-like degrees of freedom or ``topological
nature'' of the models), which in the matrix model formulation are
symmetric functions of eigenvalues of some matrix $\Lambda$:
$t_k = \frac{1}{k} {\rm tr} \Lambda^{-k}$, $k = 1,2,\ldots$ Partition
function is essentially given by the matrix integral
\be
\tau(\Lambda|W) \sim \int dX \exp \left( -{\rm tr}\int^X W(x)dx +
  {\rm tr} XW'(\Lambda)\right),
\label{GKM}
\ee
and in this particular case (\ref{combin}) can be stated explicitly
\cite{Krit}:
\be
\tau(\Lambda|W) \sim \tau_p \left( t_k + \frac{p}{k(p-k)}
{\rm res} W^{1-\frac{k}{p}}(x)dx\right)
\label{combinGKM}
\ee
The interesting property of $\tau_p$ is that it is usually a KP
$\tau$-function with $t$'s being just KP time-variables. This
observation raises a lot of questions, which can be studied in the
search for adequate generalizations of GKM and can help to make further
steps towards exact formulation and proof of the fundamental
property (\ref{combin}).

One of ideas, implied by the study of Kontsevich model, is to look
for the independent characterization of the object $\tau_p$, given
that it is a KP $\tau$-function. Usually KP $\tau$-functions are
considered as depending on two sets of variables: the ``times''
$t_k$ and the ``point of the Sato Universal Grassmannian''
$g$ \cite{SW}.
In particular our $\tau_p(t) = \tau^{(KP)}(t|g_p)$, and the
question is what is the way to characterize the point $g_p \in {\cal GR}$
without explicit reference to the matrix integral (\ref{GKM}).
Usually the point of ${\cal GR}$, as of any homogeneous manifold,
is characterized by its stability subgroup in the group
$GL(\infty)$ of symmetries of ${\cal GR}$. This is the way,
leading to the theory of ``Virasoro and $W$-like constraints''.
Alternative approach was explicitly formulated by
V.Kac and A.Schwarz \cite{KSch}. They suggested to associate the point
$g_p$ with the intersection of invariant spaces of some set of operators,
acting on a linear bundle over ${\cal GR}$, and proved that in
particular case of GKM
(at least for $p=2$) just two operators are enough to fix $g_p$
unambiguously. This formulation is inspired by the theory of
``reduced hierarchies'' and ``string equations'', as well as by
older considerations of integrable hierarchies in terms of
pseudodifferential operators.
It is an appealing approach, because it
allows to study the problem in much more generality, asking what
happens for arbitrary choice of operators etc.

Unfortunately, this seems to remain an almost untouched field, at
least we are not aware of exhaustive discussion even of the
following basic problems:

(a) What is the way to find some set of operators, associated with
{\it any} given point $g \in {\cal GR}$ and what is the way to
characterize the ambiguity of the set?

(b) What is the adequate basis in the space of all operators,
acting on Grassmannian - adequate for this kind of problems?

(c) What characterizes the minimal set (at least the number of
operators), needed to define the specific point $g \in {\cal GR}$?

(d) What - in full generality, i.e. for any $g \in {\cal GR}$ -
is the relation between the Kac-Schwarz problem of intersection
of {\it invariant} subspaces and that, characterizing $g$ as
a {\it stable} point of some subalgebra of symmetries ($GL(\infty)$)
of ${\cal GR}$ (i.e. in terms of Virasoro or $W$-like constraints)?
What (if any) is the group-theoretical interpretation of the
relevant invariant subspaces (representations)?

\centerline{$\ldots$}

(e) When does a matrix-integral representation of $\tau(t|g)$
exist (for which, if not any, $g$), and how is the number of
matrix integrations (at least) related to the set of the
corresponding Kac-Schwarz operators?\footnote{This question is
motivated by the belief that KP $\tau$-functions are usually
associated with string models with no gauge fields (``$c\leq 1$''
set is the example). In the language of matrix models this
implies the possibility to ``decouple''
in one or another way the integration over angular (unitary)
matrices, thus leaving only that over eigenvalues. If inversed,
the hypothesis would be that any KP $\tau$-function - which can
be always formulated (after a kind of Fourier transform) as an
``eigenvalue model'' - can be actually lifted to some matrix
integral where angular matrix integration can be performed
exactly.}

\centerline{$\ldots$}

In these notes we will not achieve much progress in discussion of
these problems. Our goal is modest: to show that the questions,
including (e) are, perhaps, not senseless. The need for such
demonstration does exist, since one could (and still can) simply assume,
that existence of matrix integral representation for $\tau(g)$
is an exclusive property of specific points $g = g_p$. This
suspicion could be partly supported by the failure (so far)
to find such representation for the only generalization of $g_p$'s,
which was ever discussed: for points $g_{p,q}$ (associated - in
some peculiar sense - with $(p,q)$ rather than $(p,1)$ minimal
conformal models). Here it is known only that the analogue of
the integral (\ref{GKM}) defines the duality transformation
$\tau_{p,q} \rightarrow \tau_{q,p}$ \cite{KhMa}\footnote{
$\tau_{p,q} \neq \tau_{q,p}$! The reference to minimal
conformal models could erroneously suggest that they are equal,
but the symmetry between $p$ and $q$ is
broken by coupling to $2d$ gravity, for example $\tau_{1,p} = 1$
while $\tau_{p,1} = \tau_p$, defined in (\ref{combinGKM}).}, but
no matrix integral was so far discovered to represent $\tau_{p,q}$
itself for $q > 1$.
Perhaps, however, the failure is only due to the small number of
solvable generalizations of GKM, which were studied so far, and
further work in this direction can bring a solution.
Parameters $p,q$ of the $(p,q)$-models
have an interpretation as  ``gap sizes'' in terms of the Kac-Schwarz
operator\footnote{In the language of pseudodifferential operators
the Kac-Schwarz operators are represented as polinomials in
$\frac{\partial}{\partial t_1}$. Then  $p$ and $q$ are just the orders
of these polinomials. See \cite{AVMS} for more details and references.}
(see next Section).  Before addressing the
problems of $q >1$ it can be reasonable to pay more attention to the
deformations of Kac-Schwarz operators which preserve the unit gap size,
$q=1$.

In this paper we analyze generic gap-one Kac-Schwarz
operator and show that  the corresponding $\tau(g)$
can indeed be lifted to some matrix integral, which is non-trivial
multi-matrix generalization of GKM.
We actually derive this
integral representation only in the simplest case (for some
special choice of parameters in the Kac-Schwarz problem), but
it seems to exist at least in generic {\it gap-one} situation.
This succesful experiment can probably encourage further
investigation of the whole set of above-mentioned problems.

\section{Kac-Schwarz operators}

\subsection{KP $\tau$-function in Miwa coordinates}

KP $\tau$-function is most conveniently defined as a generating
function of all the correlators in the system of free fermions
\cite{J}:
\be
\tau^{(KP)}(t | g) =
\langle 0 | e^H g | 0\rangle,
\label{tau}
\ee
where
\be\label{not}
H = \sum_{k>0} t_kJ_k, \ \ \
g = \exp \sum_{m,n}{\cal G}_{mn}\psi_m\tilde\psi_n, \nn \\
J(z) = \sum_{k = -\infty}^{+\infty} J_k z^{-k-1} = \ :\psi(z)\tilde\psi(z):
\nn \\
\psi(z) = \sum_{k = -\infty}^{+\infty} \psi_k z^{k}, \ \
\tilde\psi(z) = \sum_{k = -\infty}^{+\infty} \tilde\psi_k z^{-k-1},
\ \nn \\
\{\psi(z),\tilde\psi(z')\}_+ = \delta(z-z'),\ \ \ [J(z),J(z')]_-
= \delta'(z-z'), \nn \\
\langle 0 | \psi_k = 0 \ \ {\rm for}\ k\geq 0,  \ \
\langle 0 | \tilde\psi_k = 0 \ \ {\rm for}\ k < 0.
\ee
The one-parameter discrete sequence of $t$-variables is in fact enough
to generate any correlator of fermions, provided ``big'', not only
infinitesimal variations of $t$ are allowed\footnote{
This statement reflects nothing but the fact that
universal enveloping of the Kac-Moody
algebra $\hat G_k$ coincides with that of its Heisenberg-Cartan
subalgebra, provided $G$ is simply laced and $k=1$. In the case of KP
theory $G$ is just $U(1)$.}.
The basic formula is:
\be
\Psi(\lambda,\mu) \equiv
\frac{\langle \psi(\lambda)\tilde\psi(\mu) e^H g \rangle}
{\langle e^H g \rangle} =
\frac{{\cal X}(\lambda,\mu)\tau(t | g)}{\tau(t | g)},
\label{Psi}
\ee
where ``vertex operator'' ${\cal X}(\lambda,\mu)$ performs
the Backlund-Miwa $GL(\infty)$ transformation of $\tau$-function:
\be
{\cal X}(\lambda,\mu) = \frac{1}{\lambda - \mu}
\exp\left(\sum_{k>0} t_k(\lambda^k - \mu^k)\right)
\exp\left(\sum_{k>0} \frac{\lambda^{-k} - \mu^{-k}}{k}
\frac{\partial}{\partial t_k}\right) = \nn \\
= \frac{1}{\lambda-\mu} e^{V(\lambda)-V(\mu)}e^{\eta(\lambda)-\eta(\mu)}
= \left(e^{V(\lambda)}e^{\eta(\lambda)}\right)
\left(e^{-V(\mu)}e^{-\eta(\mu)}\right), \nn \\
V(\lambda) = \sum_{k>0} t_k\lambda^k, \ \ \
\eta(\lambda) = \sum_{k>0} \frac{\lambda^{-k}}{k}\frac{\partial}
{\partial t_k}.
\label{X-op}
\ee
Operator $e^{\eta(\{\lambda_\alpha\},\{\mu_\alpha\})} =
\prod_{\alpha = 1}^n e^{\eta(\lambda_\alpha) - \eta(\mu_\alpha)}$
shifts the time variables according to the rule
\be
t_k \ \rightarrow \ t_k\{\lambda_\alpha,\mu_\alpha\} =
t_k^{(0)} + \frac{1}{k} \sum_{\alpha = 1}^n (\lambda_\alpha^{-k} -
\mu_\alpha^{-k}).
\label{miwatr}
\ee
For $e^{\eta(\lambda,\mu)}$, entering the definition of
${\cal X}(\lambda,\mu)$ $n=1$. However, the same transformation
with $n > 1$, implying insertion of $n$ pairs of $\psi$ and $\tilde\psi$
operators under the average-sign in (\ref{tau}), can be reduced to the
$n=1$ case with the help of Wick theorem for Gaussian functional
integrals (also called Fay's identity in the theory of $\tau$-functions):
\be
\Psi(\{\lambda_\alpha\},\{\mu_\alpha\}) \equiv
\Delta(\lambda)\Delta(\mu) \frac{\prod_{\alpha =1}^n
{\cal X}(\lambda_\alpha,\mu_\alpha) \tau(t | g)}{\tau(t | g)} =
\det_{1 \leq \alpha,\beta \leq n} \Psi(\lambda_\alpha,\mu_\beta),
\label{Wick}
\ee
where
\be
\Delta(\lambda) = \prod_{\alpha > \beta}(\lambda_\alpha - \lambda_\beta) =
\det_{1 \leq \alpha,\beta \leq n} \lambda_\alpha^{\beta - 1}.
\label{Vander}
\ee

One can further consider taking $n = \infty$ in this formula.
Then one can think that all the information about time-variables
is encoded in $\lambda$-variables, while $t_0$ in (\ref{miwatr})
can be, if
necessary, absorbed into $g$. In fact, once $\lambda$'s are
allowed to be complex, the double set $\{\lambda, \mu\}$
is unnecessarily large for the purpose of fully parametrizing
the $t$-space, and one can eliminate $\mu$'s by puting them all
equal: $\mu_\alpha = \mu_*$. Usually one chooses $\mu_* = \infty$.
Then eq.(\ref{Wick}) can be regarded as a formula for $\tau$-function
itself, expressed through Miwa coordinates:
\be
\tau\{\Lambda|g\} =
\frac{\det_{1\leq \alpha,\beta \leq N}
\Psi_\alpha(\lambda_\beta)}{\Delta(\lambda)}
\label{taumiwa}
\ee
where
\be
\Psi(\lambda,\mu) = \sum_{\alpha\geq 1}\mu^{-\alpha}\Psi_\alpha(\lambda).
\ee
Since $\Psi(\lambda,\mu) = \frac{1}{\mu-\lambda}(1 + o(\lambda - \mu))$,
\be
\Psi_\alpha(\lambda) = \lambda^{\alpha - 1}(1 + o(\lambda^{-1})).
\label{norm}
\ee
The set of functions in eq.(\ref{taumiwa}) is not fixed unambiguosly
for given $g$: any linear triangular transfromations
\be
\Psi_\alpha(\lambda) \rightarrow \Psi_\alpha(\lambda) +
\sum_{1\leq\beta < \alpha} A_{\alpha\beta}\Psi_\beta(\lambda)
\label{litr}
\ee
does not change the determinant in (\ref{taumiwa}), and thus
leaves $\tau$-function the same.
This freedom can be used to bring all the functions to
``canonical'' form,\footnote{For relation between
Matrices $S_{\alpha\beta}$ in (\ref{normcan}) and
${\cal G}_{mn}$ in (\ref{not}) see \cite{Zab}. }
\be
\Psi_\alpha^{({\rm can})}(\lambda) = \lambda^{\alpha - 1} +
\sum_{\beta \geq 1}S_{\alpha\beta}\lambda^{-\beta} =
\lambda^{\alpha - 1} + o(\lambda^{-1}),
\label{normcan}
\ee
but we are not going to use this ``gauge-fixing'' below.
$\Psi_\alpha$'s can be considered as defining the coordinates
in a bundle ${\cal B}$ over
``infinite-dimensional Universal Grassmannian'' ${\cal GR}$
\cite{SW}. The point of Grassmannian $g$ is associated with
the class of equivalency
\be
{\cal W}_g = \{\Psi_1(\lambda), \Psi_2(\lambda), \ldots\}
\ee
defined modulo transformations (\ref{litr}).

\section{Formulation of the Kac-Schwarz problem}

Idea of the Kac-Schwarz approach is to consider linear operators acting
on ${\cal B}$ and look on their invariant subspaces.
An example of such operator is the ``$p$-reduction'' constraint,
\be
\lambda^p {\cal W}_g \subset {\cal W}_g,
\ee
which is known to specify the subset of ${\cal GR}$, on which
the KP $\tau$-functions are (almost) independent on all the
time variables $t_{pk}$. This constraint has a natural generalization
when $\lambda^p$ is substituted by any function $W_p(\lambda)$ of
power $p$, i.e. $W_p(\lambda) = \lambda^p\left(1 +
o(\lambda^{-1})\right)$ (it actualy does not need to be a polinomial). Then
appropriate change of time-variables brings such constraint into the standard
$p$-reduction one \cite{Krit}.  The variables on which $\tau$-function is
(almost) independent are then given by $t^{(W)}_{pk} = \frac{1}{k} {\rm tr}
(W_p(\lambda))^{-k}$.  The $W_p$-reduction constraint implies that
\be
W_p(\lambda) {\cal W}_g \subset {\cal W}_g: \nn \\
W_p(\lambda)\Psi_\alpha(\lambda) =
\Psi_{\alpha + p} + \sum_{1\leq \beta < \alpha+p}
\Omega _{\alpha\beta}\Psi_\beta(\lambda) \label{Wpal}
\ee
and it
provides an example of a ``gap-$p$'' operator,
for which ${\cal O}{\cal W}_g$ is
a subspace of codimension $p$ in ${\cal W}_g$. Coefficients $\Omega$ are at
this moment defined up to conjugation by  transformations (\ref{litr}).

This kind of constraints, however, is not enough to restrict
the point $g$ strongly enough: there remains still a large freedom
(there are plenty of KdV ($p=2$) $\tau$-functions, for example).
The idea is then to impose {\it more} constraints, associated with
some other operators, and fix $g$ more strictly as the point
associated with intersection of all the invariant subspaces in
${\cal B}$.
Therefore it is a natural problem to consider after (\ref{Wpal})
the system of two constraints:
\be
W_p(\lambda){\cal W}_g \subset {\cal W}_g, \nn \\
{\cal K}{\cal W}_g \subset {\cal W}_g
\label{constr}
\ee

Let us assume further that ${\cal K}$ is a gap-one operator. This
is a very restrictive condition. It implies
that the coordinates of the point $g$ in Grassmannian can be
represented by
\be
{\cal W}_g = \left\{ \Psi_1(\lambda), {\cal K}_1(\lambda)\Psi_1(\lambda),
\ldots \right\},\ \ \ {\rm i.e.} \ \
\Psi_\alpha(\lambda) = {\cal K}_1^{\alpha -1}\Psi_1(\lambda).
\label{gap1gr}
\ee
This condition already fixes the ``gauge freedom'' of linear triangular
transformations (\ref{litr}).
Let us agree that the other constraint is written in the form
(\ref{Wpal}) in exactly this basis, thus eliminating the ambiguity
in the choice of coefficients $\Omega$.
Then we get the basic linear equation on $\Psi_1(\lambda)$, arising from
(\ref{Wpal}) for $\alpha = 1$:
\be
\left({\cal P}_p({\cal K}_1) - W_p(\lambda)\right) \Psi_1(\lambda) = \nn \\ =
\left( {\cal K}_1^p + \sum_{1\leq \beta < \alpha+p} \Omega _{1,\beta}
{\cal K}_1^\beta - W_p(\lambda)\right) \Psi_1(\lambda) = 0,
\label{baseq}
\ee
while the rest of the equations (\ref{Wpal}) is essentially a
consistency condition for the system (\ref{constr}),
determining the coefficients $\Omega_{\alpha\beta}$ for all $\alpha > 1$.
$\tau$-function in this case is given by
\be
\tau\{\Lambda | g\} = \frac{\det_{(\alpha\beta)} {\cal K}_1^{\alpha-1}
\Psi_1(\lambda_\beta)}{\Delta(\lambda)}.
\label{tauK}
\ee

So far we did not ask anything from ${\cal K}_1$ except for being a
gap-one operator. Under this condition it is senseless to take
${\cal K}_1$ to be a function: then (\ref{baseq}) will be either true
identically (for the adequate rigid choice of $\Omega_{1,\beta}$),
giving a $\tau$-function of the factorized form,
\be
\tau\{\Lambda | g\} = \prod_{\alpha = 1}^{\infty} \Psi_1(\lambda_\alpha)
\ee
for {\it any}
\be
\Psi_1(\lambda) = 1 + o(\lambda^{-1}),
\label{noco}
\ee
or instead have only a delta-function like solution for $\Psi_1$,
which does not satisfy the normalization condition (\ref{noco}).

Generic linear operator ${\cal O}$ acting on ${\cal B}$ is
non-local. One can, however consider it as belonging to the
universal envelopping of an algebra,
generated by functions and first-order {\it differential} operators.
This can serve as excuse for selecting ${\cal K}_1$ to
be the {\it first}-order differential operator:
\be
{\cal K}_1(\lambda) = A(\lambda)\frac{\partial}{\partial\lambda} +
B(\lambda) = h(\lambda)
\left(A(\lambda)\frac{\partial}{\partial\lambda}\right)
h(\lambda)^{-1},
\nn \\
h(\lambda) = \exp\left(-\int^{\lambda} \frac{B(x)}{A(x)}dx\right).
\ee
Rescaling Baker-Ahiezer function,
\be
\Psi_1(\lambda) = \exp\left(-\int^\lambda \frac{B(x)}{A(x)}dx\right)
\Phi(\lambda) = h(\lambda)\Phi(\lambda),
\label{Resc}
\ee
we obtain equation (\ref{baseq}) in the form:
\be
\left({\cal P}_p\left(A(\lambda)\frac{\partial}{\partial\lambda}\right)
\ -\ W_p(\lambda)\right) \Phi(\lambda) = 0
\label{baseqq}
\ee
with arbitrary polinomial ${\cal P}_p$ of power $p$.

Moreover, the consistency condition
\be
[{\cal K}_1, W] = F(W)
\ee
requires that
\be
A(\lambda) = \frac{F(W)}{W'},
\ee
while the {\it unity-gap}
requirement allows $F$ to be either of power less then two in
$W$,\footnote{In some special cases, e.g. for $p=2$,
$F(W) = W^2$ is still allowed, but we do not consider such
exceptional cases here. Note also that our definition of
``power-$p$'' function implies that the coefficient in front
of the $p$-th power is unity, thus ``function of power $p$''
is not identically zero.}
$F(W) = W^\sigma(1+ o(W^{-1}))$, $\sigma \leq 1$.
The same unity-gap condition implies
that $B(\lambda)$ is of power one. It is of course non-vanishing,
this is important for (\ref{gap1gr}) to be true.
Finally one should pick up
solutions to (\ref{baseqq}) satisfying normalization
requirement (\ref{noco}).

Thus, if restricted to a system of two constraints, one defined
by a {\it function}, another - by a {\it gap-one} first-order differential
operator, the Kac-Schwarz problem actualy depends only on the
choice of the power-$p$ function $W_p$, polinomial ${\cal P}_p$
and the power-one function $F(W)$.
We shall now study this problem for two particular choices of
$F(W)$:  $F(W) = {\rm const}\ $ and $F(W) = {\rm const}\cdot W$.
In the first case for any
$W_p$ and ${\cal P}_p$ the answer is represented by
Generalized Kontsevich model,
while in the second case it is a slightly more sophisticated
model, which can be considered as peculiar average of GKM over
external matrix field (compare with \cite{char}).

\section{The case of $F(W) = c^{-1}\ $ (the GKM)}

Perform an integral transformation
\be
\Phi(\lambda) =  \int dx e^{cuW(\lambda)} f(x)
\label{ans}
\ee
and substitute it into (\ref{baseqq}):
\be
\left({\cal P}\left(c^{-1}\frac{\partial}{\partial W(\lambda)}\right) -
W(\lambda)\right) \Phi(\lambda) = \nn \\ =
\int dx e^{cxW(\lambda)} f(x) \left({\cal P}(x) - W(\lambda)\right) = 0.
\ee
This equation implies that the integrand at the r.h.s.
is total derivative w.r.to $x$, so that the integral vanishes
for appropriate choice of integration contour.
This implies in turn that
\be
f(x) = \exp\left(-c\int^x {\cal P}(x)dx \right).
\label{fP}
\ee

With our choice of integral transformation  the
powers of operator
${\cal K}_1 = h(\lambda)\frac{\partial}{c\partial W(\lambda)}
h(\lambda)^{-1}$
act on the Baker-Akhiezer function
$\Psi_1(\lambda) = h(\lambda)\Phi(\lambda)$ just by insertion
of powers of $x$ under the integral sign in (\ref{ans}):
\be
{\cal K}_1^\alpha \Psi_1(\lambda) \sim
h(\lambda)\int dx e^{cxW(\lambda)}f(x) x^\alpha.
\ee
where ``$\sim$'' sign means equivalence up to linear triangular
transformationms (\ref{litr}) which leave $\tau$-function
(\ref{taumiwa}) intact. Because of this we obtain from (\ref{tauK}):
\be
\tau\{\Lambda|g\}
= \frac{1}{\Delta(\lambda)}
\prod_{\alpha = 1}^n h(\lambda_\alpha)\int dx_\alpha f(x_\alpha)
e^{cx_\alpha W(\lambda_\alpha)} \Delta(x).
\ee
With the help of the Harish-\-Chandra-\-Itzykson-\-Zuber formula
for unitary matrix integration,
\be
\frac{1}{{\rm Vol}_{U(n)}}\int_{n\times n} [dU] e^{{\rm tr}UXU^{-1}Y} =
{\rm det}_{\alpha\beta} \frac{e^{x_\alpha y_\beta}}{\Delta(x)\Delta(y)},
\nn \\
{\rm Vol}_{U(n)} = \frac{(2\pi)^{n(n+1)/2}}{\prod_{k=1}^n k!},
\label{HCIZ}
\ee
and explicit expression (\ref{fP})
this eigenvalue integral can be rewritten as a matrix integral:
\be
\tau\{\Lambda|g\} \sim
\int_{n\times n} dX \exp\left(c\ {\rm tr}\int^X {\cal P}(x)dx
+ c\ {\rm tr} XW(\Lambda)\right).
\label{KIpre}
\ee

Following \cite{GKM} we can now change the variable $\lambda \rightarrow
\tilde\lambda(\lambda)$ so that\be
W(\lambda) = {\cal P}(\tilde\lambda).
\ee
Since both ${\cal P}$ and $W$ are functions of the same power $p$
this is allowed change of $\lambda$-variables
$\tilde\lambda = \lambda (1 + o(\lambda^{-1}))$, and the
normalization condition
\be
\Phi(\lambda) = 1 + o(\lambda^{-1})
\ee
is actually not affected.

In these new variables  (\ref{KIpre}) acquires the standard
form of Generalized Kontsevich model \cite{GKM}:
\be
\tau\{\Lambda|g\}\ \longrightarrow\ \nn \\
\tau\{\tilde\Lambda|g_{\cal P}\} \sim
\int_{n\times n} dX \exp\left(c\ {\rm tr}\int^X {\cal P}(x)dx
+ c\ {\rm tr} X{\cal P}(\tilde\Lambda)\right).
\label{KIpre'}
\ee

\section{The case of $F(W) = W$}

\subsection{Solution for $\Phi(\lambda)$}

In this case it is convenient to perform the
change of variables\footnote{See \cite{Krit} for
more detailed discussion of such changes, peculiar for the theory
of ``equivalent hierarchies''.}
$\hat\lambda = W^{1/p}(\lambda)$ at intermediate stages of calculation.
After this change the main equation (\ref{baseqq}) turns into:
\be
\left({\cal P}_p\left(c^{-1}\hat\lambda
\frac{\partial}{\partial \hat\lambda}\right) - \hat\lambda^p\right)
\Phi(\lambda) = 0
\label{baseq2}
\ee
Let us perform integral transformation
\be
\Phi(\hat\lambda) = \hat\lambda^b \int_C du f(u) e^{c\hat\lambda u}.
\label{ansa}
\ee
Note that it is different from the one used in Kontsevich case
in the previous section, since $\hat\lambda$ appears in the exponent
instead $W(\lambda) = \hat\lambda^p$: this is adequate to the new
form of the differential operator involved.
Substitution of (\ref{ansa}) into (\ref{baseq2}) gives
\be
\int du f(u) e^{c\hat\lambda u} \left(\tilde{\cal P}_p(\hat\lambda u) -
\hat\lambda^p\right) = 0,
\label{baseq3}
\ee
where $\tilde{\cal P}_p(y)$ is some power-$p$ polinomial, built
from ${\cal P}_p(x)$. It is actually simpler to say what is ${\cal P}(x)$
for a given $\tilde{\cal P}(y)$:\footnote{Direct relation is as follows.
Let $\tilde{\cal P}^{(k)}(y)$ be associated with monomial
${\cal P}(x) = x^k$. These satisfy an obvious recurrent formula:
\be
\tilde{\cal P}^{(k+1)}(y) =
\frac{1}{c}\left((cy+b)\tilde{\cal P}^{(k)}(y) +
y\frac{\partial}{\partial y} \tilde{\cal P}^{(k)}(y)\right)
\nn
\ee
and as a corollary the generating functional
\be
\tilde{\cal P}(cv|y) = \sum_{k\geq 0}^\infty
\frac{(cv)^k}{k!}\tilde{\cal P}^{(k)}(y) =
\exp\left(bv + c(e^v-1)y\right).
\nn
\ee}
\be
{\rm if}\ \ \tilde{\cal P}(y) = y^k\ \ {\rm then} \nn \\
{\cal P}^{(k)}(x) = c^{-k}(cx-b)(cx-b-1)\ldots (cx-b+1-k) = \nn \\
= c^{-k}\frac{\Gamma(cx+1-b)}{\Gamma(cx+1-b-k)}
\label{invtr}
\ee
It remains to adjust function $f(u)$ to the polinomial
$\tilde{\cal P}_p(x)$ so that the integrand in (\ref{baseq3})
becomes full derivative.\footnote{
In generic theory of matrix models one often does not need to care
about exact choice of integration contour for which the integral of total
derivative is actually vanishing. To make the model physically sensible
and even to respect the normalization conditions like (\ref{noco}) -
necessary to have the {\it standard} interpretation of
eq.(\ref{taumiwa}) for $\tau$-functions, a rather sophisticated choice
can be required, especially when exponential factors are present
in the integrand. The choice of a contour is usualy a separate
problem to be addressed independetly, see for example the next
footnote.}

We discuss the way to solve  equation (\ref{baseq2}) in full generality
in subsection 5.3 below, while now we instead consider a
simple example. In this example, instead of adjusting $f(u)$ to a
given $\tilde{\cal P}_p$ we do the opposite: choose some specifically
simple $f(u)$ and consider only $\tilde{\cal P}_p(x)$ - and thus
${\cal P}_p(x)$ - associated with it.
Namely, let us take
\be
f(u) = \frac{1}{(u^p - 1)^{r+1}}.
\label{fofu}
\ee
Then
\be
\tilde{\cal P}_p(uz) = (uz)^p - \frac{rp}{c}(uz)^{p-1}
\ee
and according to (\ref{invtr})
\be
{\cal P}_p(x) = c^{-p}(cx-b)(cx-b-1)\ldots (cx+2-b-p)
(cx + 1 - b - p - rp)
\label{speP}
\ee
We see that this is not the most general
polinomial of degree $p$: this is because we restricted
ourselves to a very special choice of function $f(u)$
in (\ref{ansa}). This choice simplifies considerably the
matrix integral representation of the $\tau$-function,
to be derived in the next subsection. After this description
we return to consideration of generic polinomials
${\cal P}_p(x)$ in (\ref{baseq2}).

It deserves saying that in order to obtain correct
asymptotics of (\ref{ansa})
as $\hat\lambda \rightarrow -\infty$, the contour $C$ in (\ref{ansa})
should be chosen to encircle all the singularities at the negative part
of the real line.\footnote{Actually for integer $r$ the whole
integral is just equal to residue at $u = 1$, while for half-integer
$r$ it is twice the integral along the real line between $u=1$ and
$u = \infty$.}
Asymptotic behaviour is actualy dictated by the vicinity of
$u = 1$ and $\Phi(\hat\lambda) \sim e^{c\hat\lambda}\hat\lambda^{b+r}$.
In order that $\Psi_1(\lambda)$ has correct asymptotics (i.e.
tends to one as $\hat\lambda \rightarrow \infty$) we need this
expression to be completely compensated by the coefficient
$h(\lambda) = \exp\left(-c\int^{\hat\lambda} \frac{B(x)dx}{x}\right)$,
distinguishing $\Psi_1(\lambda)$ from $\Phi(\lambda)$. This is an
extra restriction on parameter $b+r$ in (\ref{ansa}): it is
expressed through parameters of Kac-Schwarz operator ${\cal K}_1$.
If $B(\lambda) = \hat\lambda + a + o(\lambda^{-1}) =
W^{1/p}(\lambda) + a + o(\lambda^{-1})$, then the requirement is:
$b + r = ac$.

\subsection{Matrix integral representation}

In this section we show how solution to the
equation (\ref{baseq2}) with $f(u)$
given by (\ref{fofu}) can be lifted to a matrix integral.

The Kac-Schwarz operator ${\cal K}_1$ acts on $\Phi(\lambda)$
represented as (\ref{ansa}) by insertions of powers of $u\hat\lambda$:
\be
{\cal K}_1^{\alpha} \Psi_1(\lambda) \sim h(\lambda)
 \int du f(u) e^{c\hat\lambda u} (u\hat\lambda)^{\alpha},
\ee
where the ``$\sim$'' sign means equality modulo linear triangular
transformations (\ref{litr}) which do not change the point of
${\cal GR}$ and the value of $\tau$-function. After a change of
integration variable $v = u\hat\lambda$ the latter one is represented
as
\be
\tau\{\Lambda|g\} = \frac{1}{\Delta(\lambda)} \left(
\prod_{\beta = 1}^n \frac{h(\lambda_\beta)}{\hat\lambda_\beta}\right)
\det_{1\leq \alpha,\beta \leq n}
\left(\int dv e^v v^{\alpha-1} f(\frac{v}{\hat\lambda_\beta})\right)
 = \nn\\
= \frac{1}{\Delta(\lambda)}
\prod_{\beta = 1}^n \left(\frac{h(\lambda_\beta)}{\hat\lambda_\beta}
\int dv_\beta e^{v_\beta} f(\frac{v_\beta}{\hat\lambda_\beta})
\right) {\rm det}_{(\alpha\beta)}v_\beta^{\alpha-1}.
\ee
The last determinant at the r.h.s. is just $\Delta(v)$.

Now let us substitute our choice for $f(u)$:
\be
f(\frac{v}{\hat\lambda}) =
\frac{\hat\lambda^{p(r+1)}}{(v^p - \hat\lambda^p)^{r+1}} =
\frac{\hat\lambda^{p(r+1)}}{\Gamma(r+1)}
\int_0^\infty x^r dx e^{-x(v^p - \hat\lambda^p)}
\ee
and obtain:
\be
\tau\{\Lambda | g\} =
\frac{1}{\Delta(\lambda)}\left(\prod_\beta
\frac{\hat\lambda^{p(r+1)-1} h(\lambda_\beta)}{\Gamma(r+1)}
 \int\int x_\beta^r dx_\beta dv_\beta e^{cv_\beta}
e^{x_\beta(\hat\lambda_\beta^p - v_\beta^p)}\right) \Delta(v) = \nn \\ =
\frac{\Delta(W_p(\lambda))}{\Delta(\lambda)}
\left(\prod_\beta \frac{\hat\lambda^{p(r+1)-1}
h(\lambda_\beta)}{\Gamma(r+1)}
\int dv_\beta e^{cv_\beta} \int_0^\infty x_\beta^r  dx_\beta \right)
\times \nn \\ \times \Delta^2(x) \Delta^2(v)
\frac{\det_{(\alpha,\beta)}e^{x_\alpha W_p(\lambda_\beta)}}
{\Delta(x)\Delta(W_p(\lambda))}
\frac{\det_{(\alpha,\beta)}e^{x_\alpha v^p_\beta}}
{\Delta(x)\Delta(v^p)}
\frac{\Delta(v^p)}{\Delta(v)}
\ee
(we substituted $\hat\lambda^p = W_p(\lambda)$).
In this formula it is already easy to recognize the
integrals (\ref{HCIZ}) over angular variables, and we finally get:
\be
\tau\{\Lambda | g\} = \frac{
\left(\det W_p(\Lambda)\right)^{r+1}}{({\rm Vol}_{U(n)})^2}
\det\frac{h(W_p^{1/p}(\Lambda)}{W_p^{1/p}(\Lambda)}
{\cal S}_{R_W(p,n)}(\Lambda) \times \nn \\
\times \int_{n\times n} dV e^{\rm tr V}{\cal S}_{R(p,n)}(V)
\int_{n\times n}  dX  (\det X)^r  e^{{\rm tr} X(W_p(\Lambda)- V^p)}
\ee
The integral over $X$ is a peculiar version of Generalized
Kontsevich model (with ``zero-time'' $r$, no ``potential'' term
in the action and integration over positive definite matrices $X$ only).
Instead the $W_p(\Lambda)$ acqures a matrix-valued
but $\lambda$-independent shift by $-V^p$
and GKM partition function is further averaged over $V$
with a complicated weight. This weight includes
\be
{\cal S}_{R(p,n)}(V) \equiv \frac{\Delta(v^p)}{\Delta(v)} = \nn \\
= \prod_{\alpha>\beta}\frac{v_\alpha^p - v_\beta^p}
{v_\alpha-v_\beta} = \prod_{\alpha>\beta} (v_\alpha^{p-1} + v_\alpha^{p-1}
v_\beta + \ldots + v_\beta^{p-1})
\ee
which is already a symmetric function of $v$'s, espressible as a
polinomial in variables ${\rm tr}V^k = \sum_{1\leq\alpha\leq n}
v_\alpha^k$. The presence of this function makes integrand
explicitly $n$-dependent.\footnote{Strange as it is, this
property is not unfamiliar in the theory of matrix model:
similar phenomenon occurs in the case of the very important
Brezin-Gross-Witten model: see \cite{BGW}.}
Similarly in the preintegral factor
\be
{\cal S}_{R_W(p,n)} (\Lambda) \equiv
\frac{\Delta (W_p(\lambda))}{\Delta(\lambda)}
= \prod_{\alpha>\beta} \frac{W_p(\lambda_\alpha) - W_p(\lambda_\beta)}
{\lambda_\alpha - \lambda_\beta}.
\ee
${\cal S}_{R(p,n)}(V)$ is actually a character of the irreducible
representation of $GL(\infty)$, associated with the Young table
with rows of lenght  $l_\beta = (p-1)(n-\beta),\ 1\leq \beta \leq n$
(a ``regular ladder'' table).\footnote{This is because
the character of representation is given by the Weyl formula:
\be
\chi_{R\{l\}}(V) = \frac{\det v_\alpha^{l_\beta + n -\beta}}{\Delta(v)} =
\frac{\det v_\alpha^{l_{n-\beta+1} + \beta - 1}}{\Delta(v)}.
\nn
\ee
Taking $l_{n-\beta+1} + \beta -1 = p(\beta -1)$, we obtain
$l_\beta = (p-1)(n-\beta)$.}
${\cal S}_{R_W(p,n)}$ for non-monomial $W_p$ is a character of
{\it reducible} representation, parametrized by the coefficients of $W$.

Another pecularity of the matrix integral is specific choice of
integration contours: $X$ is required to be positive-definite
matrix, while integration contours for eigenvalues of $V$ are
going back and forth the negative real line (encircling all the
singularities which can appear on it).

\subsection{Generic polinomial}

In order to discuss generic solution of (\ref{baseq2}) let us make
a Laplace transform in the variable $\log(\hat\lambda^c)$.
After the substitution of
\be
\Phi(\lambda) = \int dk \phi(k) \hat\lambda^{ck}
\ee
eq.(\ref{baseq2}) turns into:
\be
{\cal P}_p(k)\phi(k) =
\exp\left(-\frac{p}{c}\frac{\partial}{\partial k}\right) \phi(k) =
\phi(k-\frac{p}{c}).
\label{Geq}
\ee
Further, let $s_i$, $\ i = 1,\ldots,p\ $ be the roots of ${\cal P}_p(k)
= \prod_{i=1}^p (k - s_i)$. Obviously, solution to the equation
(\ref{Geq}) is
\be
\phi(k) = \frac{(p/c)^{ck}}{\prod_{i=1}^p \Gamma(\frac{c}{p}(k - s_i)+1)}
= (p/c)^{ck}\prod_{i=1}^p \left(
\frac{\sin \frac{\pi c}{p}(k-s_i)}{\pi}
\Gamma\left(\frac{c}{p}(s_i - k)\right)\right).
\label{gens}
\ee
In the case when ${\cal P}_p(k)$ has the peculiar form
(\ref{speP}) this expression can be significantly simplified
to include only three $\Gamma$-functions:\footnote{In terms
of $\Gamma$-functions transition from (\ref{gens}) to
(\ref{pec}) involves the use of identity (which is actually
a corollary of our reasoning with the functional equation)
\be
\prod_{i=1}^p \Gamma (x + \frac{i}{p}) = (2\pi)^{\frac{p-1}{2}}
xp^{-px+\frac{1}{2}}\Gamma(px).
\nn
\ee
}
\be
{\rm for}\ \ s_i = \frac{b+i-1}{c},\ \ 1\leq i \leq p \nn \\
\phi(k) = \frac{c^{ck}}{\Gamma(ck-b+1)} = c^{ck}
\frac{\sin \pi(b-ck)}{\pi}\Gamma(b-ck, \nn \\
{\rm while\ for}\ \ s_i = \frac{b+i-1}{c},\ \ 1\leq i \leq p-1, \ \
s_p = \frac{b+p-1+pr}{c} \nn \\
\phi(k) = \frac{c^{ck}}{\Gamma(ck-b+1)}
\frac{\Gamma(\frac{ck-b-p+1}{p}+1)}{\Gamma(\frac{ck-b-p+1}{p}-r+1)} =\nn \\
= c^{ck} \frac{\sin \pi(b-ck) \sin \pi\frac{ck-b-p+1}{p}}{\pi^2}(-)^r
\times \nn \\ \times
\Gamma(b-ck)\Gamma(\frac{b+p-1-ck}{p}+r)\Gamma(\frac{ck-b-p+1}{p}+1).
\label{pec}
\ee
Substituting further integral representations for
$\Gamma$-finctions, we obtain in this peculiar case (\ref{pec}):
\be
\Phi(\lambda) = \int dk (c\hat\lambda)^{ck}
\int_0^\infty \frac{dv}{v} \int_0^\infty dy
\int_0^\infty  \frac{d\bar y}{\bar y} e^{-cv-y-\bar y} y^{r}
\times \nn \\ \times
(cv)^{b-ck} y^{(b-1-ck)/p} \bar y^{(ck-b+1)/p}
\times(sin-{\rm factors}).
\label{pec1}
\ee
while in the general situation (\ref{gens})
\be
\Phi(\hat\lambda) = \int dk (\frac{p\hat\lambda}{c})^{ck}
\prod_{i=1}^p \int_0^\infty \frac{dy_i}{y_i} e^{-y_i} y_i^{-(c/p)(k-s_i)}
\times (sin-{\rm factors}).
\label{gens1}
\ee
Integration over $k$ now gives rise to $\delta$-functions, implying
that
\be
\log\left( e^{\pm i\pi}e^{ \pm \frac{i\pi}{p}}
\frac{\bar y^{1/p}\hat\lambda}{y^{1/p} v}\right)
\ \ \longrightarrow \ \ \bar y = -y\frac{(\pm v)^p}{\hat\lambda^p}
\label{deltapec}
\ee
in the case of (\ref{pec1}) and
\be
\prod_i y_i = \frac{p}{c}\hat\lambda^p
\label{deltagens}
\ee
for (\ref{gens}).

Let us now concentrate on the case of (\ref{pec1}).
The role of $sin$-factors is reduced to ``$\pm$''  in
(\ref{deltapec}), which can be accounted for by changing the
region of integration over $v$ from the half-line $(0,\infty)$
to a contour $C$, going back and forth from infinity to $0$.
One can further change variable of integration: $y = x\hat\lambda^p$,
so that (\ref{pec1}) turns into:
\be
\Phi(\lambda) \sim
\hat\lambda^{p(r+1)+b-1}\int_C dv \int_0^\infty dx
x^{r} e^{cv -x(v^p - \hat\lambda^p)},
\ee
what is just the expression we used in \cite{char} - and thus can be of use in
the further work with
matrix models, aimed at going beyond the most simple GKM class.

\section{Acknowledgements}

The work of M.A. and P.v.M. has been partialy supported by the
NSF grant \# DMS-9203246. P.v.M. also gratefully acknowldges the
support of a NATO, a FNRS and a Francqui Foundation grant.

A.M. is indebted for the hospitality and support
to the Volterra Center at Brandeis University during the work
on this paper and to the Organizers of Buckow-95 for the
possibility to participate in the stimulating meeting.

T.S. gratefully acknowledges the hospitality of the
University of Louvain and Brandeis University.

\bigskip

{\bf APPENDIX. SYMMETRY OPERATORS ON ${\cal GR}$}

\bigskip

This Appendix describes explicitly how the
$GL(\infty)$ transformations act on Grassmannian.

Grassmannian, which is of interest in the theory of Cartanian
$\tau$-functions is the set of hyperplanes of one-half
dimension in a linear space. In order to parametrize it one
can first pick up
a ``reference'' hyperplane, by separating the
basis in the full vector space into two equal parts:
$\{e_\alpha, e_{-\alpha}\}$.
Then the basis on every other hyperplane can
be decomposed as:
\be
E_\alpha = e_\alpha + \sum_\beta S_{\alpha\beta}e_{-\beta},
\ee
or just
\be
E = e_+ + Se_-.
\nn
\ee
Infinitesimal action of $GL(\infty)$ (rotations) on the original
vector space,
\be
e_+ \rightarrow  e_+ + ae_+ + be_-, \nn \\
e_- \rightarrow  e_- + ce_+ + de_-
\ee
moves $\ E\ $ into $\ (I + a + Sc)e_+ + (S + b + Sd)e_-$. Now rotations of
the ``positive'' vectors only, $e_+ \rightarrow (I+a+Sc)^{-1}e_+$,
should be used to bring the result back into original form, so that
\be
S \rightarrow (I+a+Sc)^{-1}(S+b+Sd),
\nn
\ee
i.e. $GL(\infty)$ act on $S$ by rational (non-linear) transformation.
Infinitesimaly,
\be
\delta S = b + aS - Sd - ScS,
\ee
or
\be
\delta S_{\alpha\beta} = b_{\alpha\beta} +
\sum_\gamma a_{\alpha\gamma}S_{\gamma\beta} -
\sum_\gamma S_{\alpha\gamma}d_{\gamma\beta} -
\sum_{\gamma,\delta} S_{\alpha\gamma}c_{\gamma\delta}S_{\delta\beta}.
\label{GRtr}
\ee

After this small exercise from the theory of homogeneous spaces
let us turn to description of Grassmannian in terms of Baker-Akhiezer
functions. Infinitesimal action of $GL(\infty)$ on $\tau$-function
is in fact described by the operator ${\cal X}(\lambda,\mu)$ in
(\ref{X-op}):
\be
\delta \tau = {\cal X}(\lambda,\mu)\tau = \Psi(\lambda,\mu)\tau.
\nn
\ee
The question is now what is the action of the same
trans\-for\-mation on
$\Psi(\lambda',\mu')$. The answer can be straight\-for\-wardly
derived as follows:
\be
\delta \Psi(\lambda',\mu') =
\delta\left(\frac{{\cal X}(\lambda',\mu')\tau}{\tau}\right) =
-\frac{\delta\tau {\cal X}(\lambda',\mu')\tau}{\tau^2} +
\frac{{\cal X}(\lambda',\mu')\delta\tau}{\tau} = \nn \\ =
-\Psi(\lambda',\mu')\Psi(\lambda,\mu) +
\frac{1}{\tau}{\cal X}(\lambda',\mu')\left(\Psi(\lambda,\mu)\tau\right).
\ee
The second item at the r.h.s. is just
\be
\Psi(\{\lambda',\lambda\},\{\mu',\mu\}) =
\Psi(\lambda',\mu')\Psi(\lambda,\mu) -
\Psi(\lambda',\mu)\Psi(\lambda,\mu'),
\ee
where Wick theorem was applied for $n=2$, and we finally obtain:\footnote{
One could use the already known result from ref.\cite{ASvM}, saying that
\be
\frac{\delta\Psi_(\lambda',\mu')}{\Psi_(\lambda',\mu')}
= \left(e^{\eta(\lambda',\mu')}- 1\right)\Psi(\lambda,\mu).
\nn
\ee
Then, using the Wick theorem in the form
\be
e^{\eta(\lambda',\mu')}\Psi(\lambda,\mu)
= \frac{(\lambda-\lambda')(\mu-\mu')}{(\lambda - \mu')(\mu - \lambda')}
\frac{e^{V(\lambda)-V(\mu)} e^{\eta(\{\lambda',\lambda\},\{\mu',\mu\})}
 \tau(t|g)}
{e^{\eta(\lambda',\mu')} \tau(t|g)} = \nn \\ =
\frac{\Psi(\{\lambda',\lambda\},\{\mu',\mu\})}
{\Psi(\lambda',\mu')} =
\frac{\Psi(\lambda',\mu')\Psi(\lambda,\mu) -
\Psi(\lambda',\mu)\Psi(\lambda,\mu')}{\Psi(\lambda',\mu')},
\nn
\ee
one obtains:
\be
\left( e^{\eta(\lambda',\mu')} - 1 \right)\Psi(\lambda,\mu) =
-\frac{\Psi(\lambda',\mu)\Psi(\lambda,\mu')}{\Psi(\lambda',\mu')}.
\nn
\ee
Residue at the pole $\frac{1}{\lambda - \mu}$ in
$\Psi(\lambda,\mu)$ at the l.h.s does not depend on time-variables
and thus the pole is annihilated by operator in the brackets.
This operator vanishes as $\lambda' = \mu'$, in accordance
with appearence of the pole of $\Psi(\lambda',\mu') =
\frac{1}{\lambda'-\mu'}(1 + o(\lambda'-\mu'))$ in denominator.
As usual, the action of such shift operator
is singular when $\lambda' = \mu$ or $\mu' = \lambda$, while
the zeroes at $\lambda' = \lambda$ and $\mu' = \mu$ do not
appear at the r.h.s. because ``$1$'' is subtracted from the
shift operator.}
\be
\delta\Psi(\lambda',\mu') =
-\Psi(\lambda',\mu)\Psi(\lambda,\mu').
\label{deltaPsi}
\ee
Expanding this relation in inverse powers of $\mu'$, we get:
\be
\delta\Psi_\alpha(\lambda') = -\Psi(\lambda',\mu)\Psi_\alpha(\lambda).
\ee
This can be now compared with eq.(\ref{GRtr}), where
\be
e_\alpha  \sim  \lambda^\alpha, \ \ \ e_{-\alpha} \sim \lambda^{-\alpha},
\nn \\
E_\alpha \sim \Psi_\alpha(\lambda).
\nn
\ee
Substitution of
\be
\Psi(\lambda,\mu) = \sum_{\alpha,\beta}\frac{\lambda^\beta}{\mu^\alpha}
(\delta_{\beta,\alpha-1} + S_{\alpha\beta})
\nn
\ee
into (\ref{deltaPsi}) gives:
\be
\delta S_{\alpha\beta} =
(\lambda^{\alpha-1} + \sum_\gamma S_{\alpha\gamma}\lambda^\gamma)
(\mu^{-\beta -1} + \sum_\gamma \mu^{-\gamma}S_{\gamma\beta}) = \nn \\
= \frac{\lambda^{\alpha-1}}{\mu^{\beta+1}} +
\sum_\gamma \frac{\lambda^{\alpha-1}}{\mu^{\gamma}} S_{\gamma\beta} +
\sum_\gamma S_{\alpha\gamma} \frac{\lambda^\gamma}{\mu^{\beta +1}}
+ \sum_{\gamma,\delta} S_{\alpha\gamma} \frac{\lambda^\gamma}
{\mu^{\delta}}S_{\delta\beta},
\ee
in accordance with the general rule (\ref{GRtr}).

\end{document}